\documentclass[a4paper]{elsarticle}
\biboptions{sort&compress}
\usepackage{lineno,hyperref}
\hypersetup{colorlinks=true}

\usepackage[utf8]{inputenc}
\usepackage[official]{eurosym}

\usepackage{amsmath}
\usepackage{IEEEtrantools}
\usepackage{mathtools}
\usepackage{siunitx}
\usepackage{tabularx}
\newcolumntype{Y}{>{\raggedleft\arraybackslash}X}

\usepackage{threeparttable}
\usepackage{enumitem}

\usepackage{comment}

\usepackage{graphicx}
\usepackage{floatrow}
\graphicspath{{graphics/}}

\usepackage{color}

\usepackage{booktabs}

\renewcommand{\d}{\mathop{}\!\mathrm{d}}

\journal{Energy Strategy Reviews}

\begin{document}

\begin{frontmatter}

\title{PyPSA-Eur: An Open Optimisation Model of the European Transmission System\tnoteref{t1}}
\tnotetext[t1]{
  \textcopyright { 2018,
    \href{https://creativecommons.org/licenses/by-nc-nd/4.0/}{CC BY-NC-ND 4.0},
    eprint of
    \href{https://doi.org/10.1016/j.esr.2018.08.012}{doi:10.1016/j.esr.2018.08.012}
  }
}

\author[iai,fias]{Jonas~Hörsch\corref{corrauthor}}
\cortext[mycorrespondingauthor]{Corresponding author}
\ead{jonas.hoersch@kit.edu}

\author[fias]{Fabian~Hofmann}
\author[fias]{David~Schlachtberger}
\author[iai,fias]{Tom~Brown}

\address[iai]{Institute for Automation and Applied Informatics, Karlsruhe Institute of Technology, 76344~Eggenstein-Leopoldshafen, Germany}
\address[fias]{Frankfurt Institute for Advanced Studies, 60438~Frankfurt~am~Main, Germany}

\begin{abstract}
PyPSA-Eur, the first open model dataset of the European power system at the transmission network level to cover the full ENTSO-E area, is presented. It contains \num{6001} lines (alternating current lines at and above \SI{220}{kV} voltage level and all high voltage direct current lines), \num{3657} substations, a new open database of conventional power plants, time series for electrical demand and variable renewable generator availability, and geographic potentials for the expansion of wind and solar power. The model is suitable both for operational studies and generation and transmission expansion planning studies. The continental scope and highly resolved spatial scale enables a proper description of the long-range smoothing effects for renewable power generation and their varying resource availability. The restriction to freely available and open data encourages the open exchange of model data developments and eases the comparison of model results. A further novelty of the dataset is the publication of the full, automated software pipeline to assemble the load-flow-ready model from the original datasets, which enables easy replacement and improvement of the individual parts.  This paper focuses on the description of the network topology, the compilation of a European power plant database and a top-down load time-series regionalisation. It summarises the derivation of renewable wind and solar availability time-series from re-analysis weather datasets and the estimation of renewable capacity potentials restricted by land-use. Finally, validations of the dataset are presented, including a new methodology to compare geo-referenced network datasets to one another.
\end{abstract}


\end{frontmatter}

\section{Introduction}

The energy system in Europe is undergoing a far-reaching transformation on multiple fronts: generation from variable renewable energy sources, such as wind and solar power, is growing due to the imperative of tackling climate change; electricity provision has been unbundled and liberalised, raising complex challenges for the efficient design and regulation of electricity markets; the need to decarbonise heating and transport is driving electrification of these sectors; and finally energy markets are being integrated across the continent~\cite{energyunion}.

To study this transformation, accurate modelling of the transmission grid is required. The need to take account of international electricity trading and the possibility of smoothing variable renewable feed-in over large distances (wind generation has a typical correlation length of around \SI{600}{km}~\cite{1748-9326-10-4-044004}) mean that models should have a continental scope. At the same time, high spatial detail is required, since national grid bottlenecks are already hindering the uptake of renewable energy today~\cite{BNetzA2016}, and given persistent public acceptance problems facing new transmission projects~\cite{Battaglinietal2012}, severe grid bottlenecks will remain a feature of the energy system for decades to come.

Currently there is no openly-available model of the full European transmission network with which researchers can investigate and compare different approaches to the energy transformation.  The transmission grid dataset provided by the European Network of Transmission System Operators for Electricity (ENTSO-E) for the 2016 Ten Year Network Development Plan (TYNDP)~\cite{STUM} is rendered unusable by restrictive licensing, the exclusion of Finland, Norway and Sweden, and a lack of geographical localisation of the represented substations. The lack of geo-data means that the crucial weather system correlations and dynamics cannot be mapped onto the network. In 2005 in \cite{Zhou2005} an openly-available model of the continental European transmission network (i.e. excluding the UK, Ireland, Scandinavia and the Baltic states) was presented using a manual matching of buses and lines to the raster graphic of the ENTSO-E map, along with an open power plant database based on the Global Energy Observatory \cite{geo}; this model was updated to the network of 2009 in \cite{Hutc13}. Apart from not covering the full ENTSO-E area, this dataset has the problem that much of the data was extracted manually, which is potentially error-prone and hard to repeat as new data becomes available, the buses are missing geo-coordinates and the power plant database is incomplete. In \cite{Jensen2017,jensen_2015_35177} geo-coordinates and data for wind and solar plants were added to the dataset.  Open datasets based on OpenStreetMap~\cite{OSM}, such as the SciGRID network~\cite{SciGRIDv0.2} and the osmTGmod~\cite{osmtgmod-github} network, are of high quality in Germany, where data is well organised, but are not yet accurate for the rest of Europe. Similarly the open electricity model provided by the German Institute for Economic Research (DIW), ELMOD-DE~\cite{ELMOD-DE}, only covers Germany.


In this paper we present a model of the European power system at the transmission network level which remedies these many deficiencies: it is not only open but also contains a high level of detail for the full ENTSO-E area. Grid data is provided by an automatic extraction of the ENTSO-E grid map; a power plant database is presented using a sophisticated algorithm that matches records from a wide range of available sources and includes geo-data; other data, such as time series for electrical load and wind, solar and hydro-electric availability, and geographic potentials for the expansion of wind and solar power, are also described.  A new technique for comparing network datasets is presented and used to validate the grid data.  The dataset and all code used to generate it from the raw data are available online \cite{pypsa-eur-zenodo,pypsa-eur-github}, as a model for the Python for Power System Analysis (PyPSA) framework version 0.13.1~\cite{pypsa,pypsa-0.13.1}.

In Section \ref{sec:methods} the data sources and processing methods are presented; the data is validated in Section \ref{sec:validation}; limitations of the dataset are discussed in Section \ref{sec:limitations}; conclusions are drawn in Section \ref{sec:conclusions}.

\section{Data sources and methods}
\label{sec:methods}

\subsection{Network topology}

The network topology and geography of substations and transmission lines have been extracted from the geographical vector data of the online ENTSO-E Interactive Map~\cite{interactive} by the GridKit toolkit~\cite{wiegmans_2016_47263}. We did not use the published extract at~\cite{wiegmans_2016_55853}, since several errors like inadvertently duplicated alternating current (AC) lines and missing transformers and lines between substations of short distances below \SI{1}{km} have been identified.

Instead we extended the GridKit toolkit\footnote{The modified toolkit has been published at \url{https://github.com/PyPSA/GridKit/}.}:
\begin{enumerate}
\item A python script was added to stitch the vector tiles from the ENTSO-E map according to the identifier attribute `oid' before importing the line structures into GridKit, since the toolkit previously often mislabeled the overlapping parts of the same line as two separate circuits.
\item The tolerance for connecting dangling high-voltage direct current (HVDC) lines and their converter stations with the next high-voltage alternating current (HVAC) substations has been increased and the new connections have been manually verified.
\item AC lines carrying circuits of several voltage levels had to be split by inspecting the descriptive text tag and are split into several lines.
\item The columns and format of the extracted CSV lines have been aligned closer with PyPSA to simplify the subsequent import.
\end{enumerate}

The electrical parameters are derived by assuming the standard AC line types in Table~\ref{tab:linetypes} for the length and number of circuits. The DC line capacities are assigned from the table in \cite{wiki:list_of_hvdc_projects}. No transformer information is contained in the map, so a single transformer of capacity \SI{2}{GW} (i.e. equivalent to four 500~MW transformers) is placed between buses of different voltage levels at the same location, with a reactance of 0.1 per unit. The transformer capacity assumption is on the high side to avoid introducing constraints where none exist in reality.

\begin{table}
  \caption{Standard line types for overhead AC lines~\cite{oeding2011-book}}
  \label{tab:linetypes}
\begin{tabularx}{\textwidth}{YSSSSSS}
\toprule
 \multicolumn{1}{l}{Volt.}  & \multicolumn{1}{l}{Wires} &  \multicolumn{1}{l}{Series}  &  \multicolumn{1}{l}{Series ind.} & \multicolumn{1}{l}{Shunt}       &  \multicolumn{1}{l}{Current}   &  \multicolumn{1}{l}{App. power}\\
 \multicolumn{1}{l}{level}    &            &  \multicolumn{1}{l}{resist.} &  \multicolumn{1}{l}{reactance}   &  \multicolumn{1}{l}{capacit.} &  \multicolumn{1}{l}{therm. limit} &  \multicolumn{1}{l}{therm. limit} \\
 \multicolumn{1}{l}{\small(kV)}     &            &  \multicolumn{1}{l}{\small($\Omega$/km)}   &  \multicolumn{1}{l}{\small($\Omega$/km)}    &  \multicolumn{1}{l}{\small(nF/km)}     &  \multicolumn{1}{l}{\small(A)}    &  \multicolumn{1}{l}{\small(MVA)} \\
\midrule
  220 & 2 & 0.06 & 0.301 & 12.5 & 1290 & 492\\
300 & 3 & 0.04 & 0.265 & 13.2 & 1935 & 1005\\
380 & 4 & 0.03 & 0.246 & 13.8 & 2580 & 1698\\
\bottomrule
\end{tabularx}
\end{table}

\begin{figure}
\floatbox[{\capbeside\thisfloatsetup{capbesideposition={left,bottom},capbesidewidth=0.3\linewidth}}]{figure}[\FBwidth]
{\caption{Transmission network model (includes optional lines that are planned and under construction).}\label{fig:europe-map}}
{\includegraphics[width=0.6\textwidth]{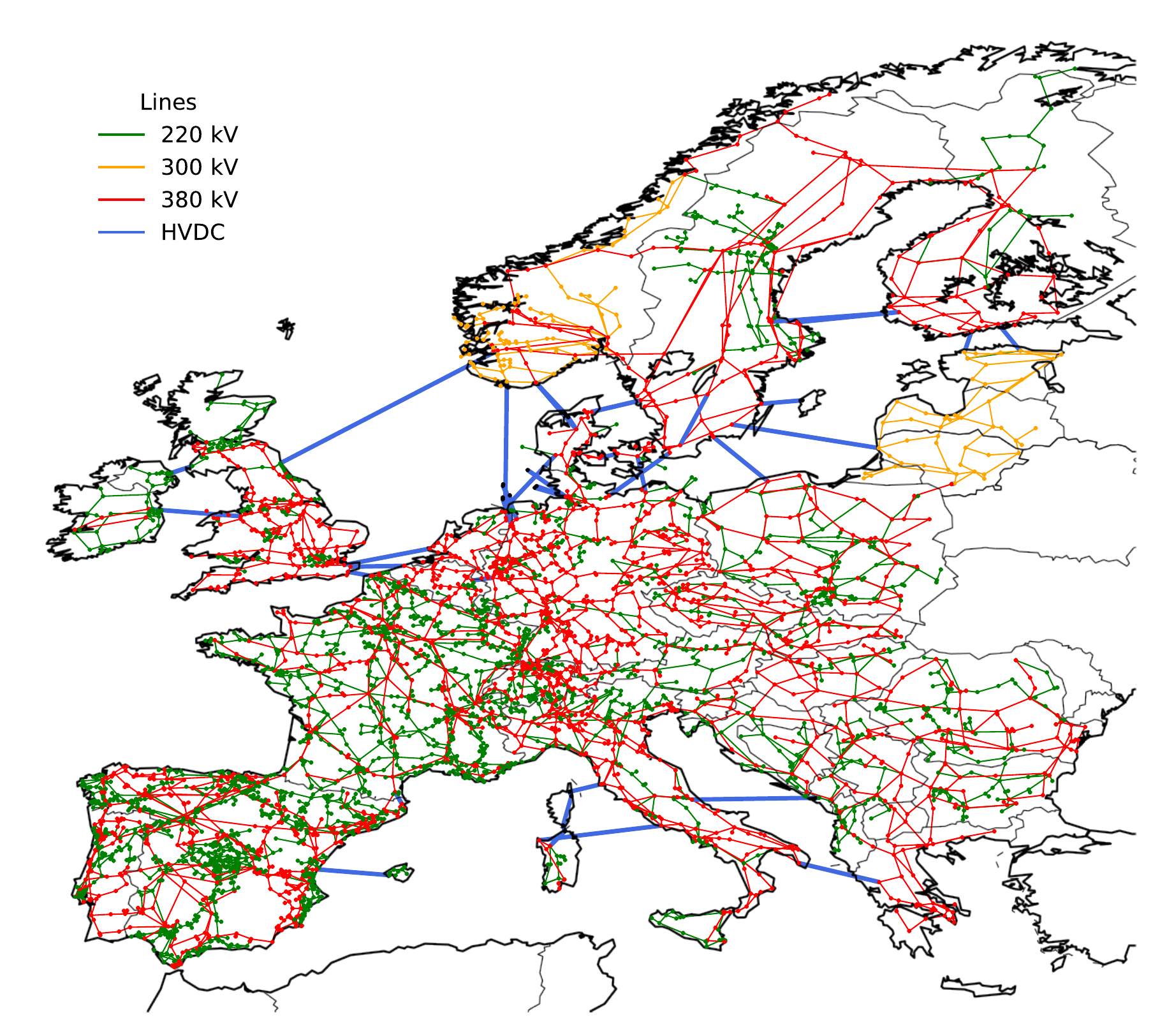}}
\end{figure}

The restriction to buses and transmission lines of the voltage levels \SIlist{220;300;380}{kV} in the landmass or exclusive economic zones of the European countries and the removal of \num{23} disconnected stub sub-networks (of less than \num{4} buses) produces the transmission network in Figure \ref{fig:europe-map} of all current transmission lines plus several ones which are already under or close to construction (these are marked in the dataset). In total the model contains \num{6001} HVAC lines with a volume of \SI{345.7}{TW.km} (of which \SI{17}{TW.km} are still under construction), \num{46} HVDC lines with a volume of \SI{6.2}{TW.km} (of which \SI{2.3}{TW.km} are still under construction). The buses are composed of \num{3657} substations and \num{1320} auxiliary buses, like joints and power plants.

The countries are partitioned into Voronoi cells as catchment areas, each of which is assumed to be connected to the substation by lower voltage network layers. These Voronoi cells are used to link power plant capacities and determine feed-in by potential renewable energy generation, as well as the share of demand drawn at the substation.

\subsection{Conventional power plants}

Official sources often only report on country-wide capacity totals keyed by fuel-type and year like the Eurostat nrg\_113a database~\cite{eurostat-nrg_113a}, the ENTSO-E net generation capacity~\cite{entsoe-netgencaps} or the ENTSO-E Scenario Outlook and Adequacy Forecast (SO\&AF)~\cite{entsoe-soaf,opsd-natgencaps}, while only seven countries\footnote{BE, DE, FR, HU, IE, IT, LT as listed by the Open Power System Data project at \url{http://open-power-system-data.org/data-sources\#23_National_sources}} have official power plant lists collected and standardised by the Open Power System Data (OPSD) project~\cite{opsd-conventionalpowerplants}.

This gap has been gradually closing since ENTSO-E started maintaining a power plant list (ENTSO-E PPL) on their Transparency Platform~\cite{entsoe-transparency}.  Unfortunately, it is still far from complete, for instance even after excluding solar and wind generators, the total capacity represented in Germany amounts only to about \SI{57}{GW}, while the SO\&AF reports \SI{111}{GW}, \SI{107}{GW} of which are also covered as operational in the German BNetzA Kraftwerksliste~\cite{bnetza}.

The powerplantmatching (PPM) tool and database~\cite{ppm-github} we present in this section achieves good coverage by (1) standardising the records of several freely available databases, (2) linking them using a deduplication and record linkage application and (3) reducing the connected claims about fuel type, technology, capacity and location to the most likely ones.

PPM incorporates several power plant databases that are either published under free licenses allowing redistribution and reuse or are at least freely accessible. In the order of approximate reliability, there are OPSD~\cite{opsd-conventionalpowerplants}, ENTSO-E PPL~\cite{entsoe-transparency}, DOE Energy Storage Exchange~\cite{doestoragedb}, Global Energy Observatory (GEO)~\cite{geo}, Carbon Monitoring for Action (CARMA) from 2009~\cite{wheeler2008-carma,ummel2012-carmav3}, DOE Energy Storage Exchange~\cite{doestoragedb} (ESE) and the Global Power Plant Database by the World Resource Institute (GPD)~\cite{wripowerwatch}. All of them are brought into the standardised tabular structure outlined in Table~\ref{tab:ppm-structure} by explicit maps between the various naming schemes and additional heuristics identifying common fuel-type or technology keywords like \emph{lignite} or \emph{CHP} in the \emph{Name} column. Furthermore, the \emph{Name} column is cleaned by removing frequently occurring tokens, \emph{power plant} or block numbers, for instance.

\begin{table}
  \centering
  \renewcommand{\tabularxcolumn}[1]{p{#1}}
\begin{tabularx}{\textwidth}{l X}
\toprule
  Column & Argument \\
  \midrule
  Name & Power plant name \\
  Fueltype         & \{Bioenergy, Geothermal, Hard Coal, Hydro, Lignite, Nuclear, Natural Gas, Oil, Solar, Wind, Other \}  \\
  Technology & \{CCGT, OCGT, Steam Turbine, Combustion Engine,
               Run-Of-River, Pumped Storage, Reservoir \} \\
  Set & \{PP, CHP\} \\
  Capacity  & Generation capacity in MW \\
  lat/lon & Latitude and Longitude  \\
  Country & \{EU-27 + CH + NO (+ UK) minus Cyprus and Malta\} \\
  YearCommissioned & Commissioning year \\
  File & Source file of the data record \\
  projectID & Identifier of the power plant in the original source file \\
  \bottomrule
\end{tabularx}
\caption{Standardised data structure for the power plant databases.}
\label{tab:ppm-structure}
\end{table}

Since OPSD, ENTSO-E PPL and ESE report individual power plant units for at least some power plants, in a first step we use the deduplication mode of the java application Duke to determine units of the same power plant. Duke~\cite{duke} is a free software extension of the search engine library Lucene that determines probabilities whether pairs of records (of the same or different tables) refer to the same entity. It computes conditional probabilities $p^{i,j}_{c} := P(M^{i,j} | x_{c}^{i,j})$ for the event $M_{i,j} := \text{``records i and j match''}$ given the data $x^{i,j}_{c}$ in column $c$ of these records from mostly character-based similarity metrics like the \emph{Jaro distance} or the \emph{Q-gram distance}~\cite{elmagarmid2007} skewed into a configurable interval and combines them into an overall matching probability as
\begin{linenomath}
\begin{equation}
  p_{i,j} := P(M^{i,j} | \cap_{c} x_{c}^{i,j}) = \frac{\prod_{c} p_{c}^{i,j}}{\prod_{c} p_{c}^{i,j} + \prod_{c} (1-p_{c}^{i,j})}~.
  \label{eq:naive-bayes}
\end{equation}
\end{linenomath}
This formula, a simplified variant of Naive Bayesian Classification, can be derived from the Bayes Theorem under the assumptions of pairwise conditional independence of the $x_{c}^{i,j}$ and unbiased prior probabilities whether two records match or do not, i.e. $P(M^{i,j}) = P((M^{i,j})^C) = 0.5$. The former assumption underlies all naive bayesian classifiers and ignores for instance the correlation between technology and capacity (run-of-the-river turbines are typical small (\SI{20}{MW}), whereas nuclear power plants are typically large, with a median capacity of \SI{2}{GW}). The latter assumption means literally that any two power plant entries from two different datasets have a prior probability of 50\% to refer to the same power plant, while the real probability is less than $\frac{N}{N^2} = \frac{1}{N}$, seen from the comparison of two identical datasets of length $N$. To include such a more realistic prior assumption into the matching process would require changing several internals of the Duke library and is out of the scope of this work. Nevertheless, the model has already been successfully applied in practice~\cite{domingos1997, androutsopoulos2000}.

For the aggregation of power plant units, Duke is configured to use the metrics and intervals described in Table~\ref{tab:ppm-comparators} to return the probabilities $p_{i,j} > 0.985$ between likely pairs $i$ and $j$. In the power plant matching tool of the authors, these are used as edges in a directed graph of records and the cliques of this graph\footnote{A clique in a directed graph is a subset of the nodes such that every two distinct nodes are adjacent.} are aggregated as power plants. Note the low end of the interval for measuring the similarity of the fuel-type chosen to prevent merging units with different fuel-types into the same power plant.

\begin{table}
  \centering
  \renewcommand{\tabularxcolumn}[1]{p{#1}}
\begin{tabularx}{\textwidth}{X l S S l S S}
\toprule
Column & \multicolumn{3}{l}{Deduplication} & \multicolumn{3}{l}{Record linkage} \\
& Comparator & low & high           & Comparator & low & high \\
\midrule
Name        & JaroToken & 0.09 & 0.99  & JaroToken & 0.09 & 0.99 \\
Fueltype    & QGram     & 0.09 & 0.65  & QGram & 0.09 & 0.7 \\
Country     & QGram     & 0.01 & 0.51  & QGram & 0.0 & 0.53 \\
Capacity    & Numeric   & 0.49 & 0.51  & Numeric & 0.1 & 0.75 \\
Geoposition & Geo       & 0.05 & 0.55  & Geo & 0.1 & 0.8\\
\bottomrule
\end{tabularx}
\caption{Duke comparison metrics and intervals for aggregation of power plant units (deduplication) and linking different power plant tables (record linkage). JaroToken breaks the full string into several tokens, evaluates the Jaro Winkler distance metric for each and returns the compound Jaccard index \cite{elmagarmid2007}. These parameters have been chosen by hand and plausibility, while instead they  should be tuned for a representative subset to an ideal match by Duke's Genetic algorithm.}
\label{tab:ppm-comparators}
\end{table}

For linking the six databases, PPM runs Duke in Record linkage mode on every pair of databases and determines the most likely links above the threshold of \num{0.985}. These links are joined to chains by collecting the records across all databases that match to the same plant in any database. The chains are reduced by keeping only the longest chains, until they are consistent, i.e. each power plant appears only in at most one chain. This could likely be improved by joining chains recursively, while keeping track of the chain probability based on a variant of Eq.~(\ref{eq:naive-bayes}) at the expense of not being able to rely on the fast pandas routines any more.

For the remaining chains the power plant information is aggregated by taking the most frequent \emph{Fueltype}, a comma separated list of the \emph{Technology}(-ies), the mean \emph{lat/lon} and the median \emph{Capacity}. The latter ensures that the shutdown or addition of a block of a power plant which is not yet reflected in a minority of databases does not distort the final capacity.

The compound dataset, at the time of writing, contains 3501 power plants with a total capacity of \SI{663}{GW}. Less than a third of these are represented in \num{3} or more sources, but still account for about two third of the capacity. \num{2584} small power plants with an average capacity of about \SI{83}{MW} appear in only two databases. There are a further 2788 power plants with \SI{18.2}{GW} capacity in the OPSD dataset unmatched by the other free datasets and exclusively compiled from official sources. After including these power plants the mean absolute error from the SO\&AF country-wise capacity is at 12\% of the average capacity and below a 27\% deviation in each single country except for Bulgaria and Lithuania. Refer to the companion paper for a more detailed comparison of the free dataset with the proprietary World Electric Power Plants dataset~\cite{gotzens2017}.

\subsection{Hydro-electric generation}

Existing hydroelectric capacities ensue from the same matching process as the conventional power plants, particularly based on the sources ESE and ENTSO-E. The capacities are categorised into run-of-river, reservoir and pumped storage. Reservoir and pumped storage have energy storage capacities that are estimated by distributing the country-aggregated energy storage capacities reported by~\cite{kies2016,pfluger2011} in proportion to power capacity. Run-of-river as well as reservoir hydro capacities receive an hourly-resolved in-flow of energy. Extensions to the current hydro capacities are not considered.

Renewable generation time series like hydro-electric in-flow, wind are derived from the re-analysis weather dataset ERA5 by the European Centre for Medium-Range Weather Forecasts (ECMWF)~\cite{era5}. It provides wind speeds, irradiation, surface-roughness, temperature and run-off in hourly resolution since 2008 on a on a $\ang{0.28} \times \ang{0.28}$ spatial raster ($x \in \mathcal{X}$).

The simplified in-flow time series is generated as in~\cite{kies2016,Schlachtberger2017} by aggregating the total potential energy at height $h_x$ relative to ocean level of the run-off data $\mathcal{R}_x$ in each country $c$ by
\begin{linenomath}
\begin{equation}
  G^H_{c}(t) = \mathcal{N} \sum_{x \in \mathcal{X}(c)} h_x \mathcal{R}_x(t)
\end{equation}
\end{linenomath}
where $\mathcal{N}$ is chosen so that $\int_{t} G^H_{c}(t) \,\mathrm{d}t$ matches the EIA annual hydroelectricity generation~\cite{EIA-hydro-netgen}. The in-flow is distributed to all run-of-river and reservoir capacities in proportion to their power capacity.

\subsection{Wind generation}
\label{sec:wind}

Adapting the methodology in~\cite{REatlas} to the newer dataset, the wind speeds at \SI{100}{m} above ground $u^{\SI{100}{m}}_x(t)$ are extrapolated to turbine hub-height $h$ using the surface roughness $z^0_x$ with the logarithmic law
\begin{linenomath}
\begin{equation}
  u^h_x(t) = u^{\SI{100}{m}}_x(t) \frac{\ln \left(h / z^0_x\right)}{\ln \left(\SI{100}{m} / z^0_x\right)}~.
\end{equation}
\end{linenomath}

The capacity factor of each raster cell $x$ for a wind turbine with powercurve $P_{w}(u)$ and generator capacity $P^{max}_w$ is determined as
\begin{linenomath}
\begin{equation}
  c_{x,w} = \frac{\left\langle\,P_{w}(u^h_x(t))\,\right\rangle_t}{P^{max}_w}
\end{equation}
\end{linenomath}
and together with the usable area $A_{x,w}$ the maximally installable wind generation capacity $G^{max}_{x,w} = 0.3 \cdot \SI{10}{MW/km^2} \cdot A_{x,w}$ is calculated, where \SI{10}{MW/km^2} is the technical potential density~\cite{Scholz} and \num{0.3} arises out of considering competing land use and issues of public acceptance.

The usable area is restricted by the following constraints: Onshore wind can only be built in land use types of the CORINE Land Cover database~\cite{corine2012} associated to \emph{Agricultural areas} and \emph{Forest and semi natural areas} and furthermore a minimum distance of \SI{1000}{m} from \emph{Urban fabric} and \emph{Industrial, commercial and transport units} must be respected. Offshore wind can only be constructed in water depths up to \SI{50}{m}. Additionally, all nature reserves and restricted areas listed in the Natura2000 database~\cite{natura2000} are excluded. The wind generation potential in Germany is shown in Figure~\ref{fig:potentials}.

Each Voronoi cell $V$ of a substation covers multiple cells of the re-analysis weather grid, as described by the indicatormatrix $\mathcal{I}_{V,x} = \mathrm{area} (V \cap x) / \mathrm{area}\,x$, and we distribute the wind turbine capacity according to a normed capacity layout
\begin{linenomath}
\begin{equation}
  G^{p.u.}_{V,x,w} = \frac{\mathcal{I}_{V,x} c_{x,w} G^{max}_{x,w}}{\sum_x \left(\cdot\right)}
  \label{eq:wind-caplayout}
\end{equation}
\end{linenomath}
which prefers cells $x$ with high capacity factor $c_{x,w}$ and high maximally installable capacity $G^{max}_{x,w}$. The wind generation availability time-series at a substation with Voronoi cell $V$ is, thus,
\begin{linenomath}
\begin{equation}
  \bar g_{V,w}(t) =  G_{V,w} \frac{\sum_x G^{p.u.}_{V,x,w} P_w(u^h_x(t))}{P^{max}_w}
  \label{eq:wind-gen}
\end{equation}
\end{linenomath}
for an installed capacity $G_{V,w}$. This capacity is expandable until reaching $G^{max}_{x,w}$ in any grid cell up to
\begin{linenomath}
\begin{equation}
  G^{max}_{V,w} = \min_{\{x | \mathcal{I}_{V,x} > 0\}}  \frac{\mathcal{I}_{V,x} G^{max}_{x,w}}{G^{p.u.}_{V,x,w}}~.
  \label{eq:wind-capcap}
\end{equation}
\end{linenomath}

The power curve of the turbine Vestas V112 with a turbine capacity of \SI{3}{MW} and a hub height \SI{80}{m} is used to generate the onshore wind time-series and the NREL Reference Turbine with \SI{5}{MW} at \SI{90}{m} is used for the offshore wind time-series. The accuracy of the wind generation time-series are improved to account for effects of spatial wind speed variations within a grid cell by smoothing the power curves with a Gaussian kernel as
\begin{linenomath}
\begin{equation}
  P_w(u) = \eta \int_0^{\infty} P_0(u'') \frac1{\sqrt{2\pi\sigma_0^2}}\,\mathrm{e}^{- \frac{(u - u' +  \Delta u)^2}{2\sigma_0^2}} \d u'~,
\end{equation}
\end{linenomath}
where $\eta = \num{0.95}$, $\Delta u = \SI{1.27}{m/s}$ and $\sigma_0 = \SI{2.29}{m/s}$ are the optimal parameters minimising the error between the re-analysis-based time-series and a year of Danish wind feed-in~\cite{REatlas}. A study comparing the wind generation time-series based on the re-analysis MERRA-2 dataset for a 20 year period to the per-country wind feed-in and several wind park generation measurements found non-negligible discrepancies of the optimal bias correction parameters between different countries~\cite{staffel2016}. They will be incorporated in a future version of the presented model.

\subsection{Photovoltaic generation}
\label{sec:PV}

\begin{figure}
  \caption{Wind (l.) and solar (r.) potential power generation after landuse restrictions for weather grid cells in Germany. The generation of all grid cells in a Voronoi cell (also shown in black) is fed into the central substation.}
  \label{fig:potentials}
\includegraphics[width=0.48\textwidth]{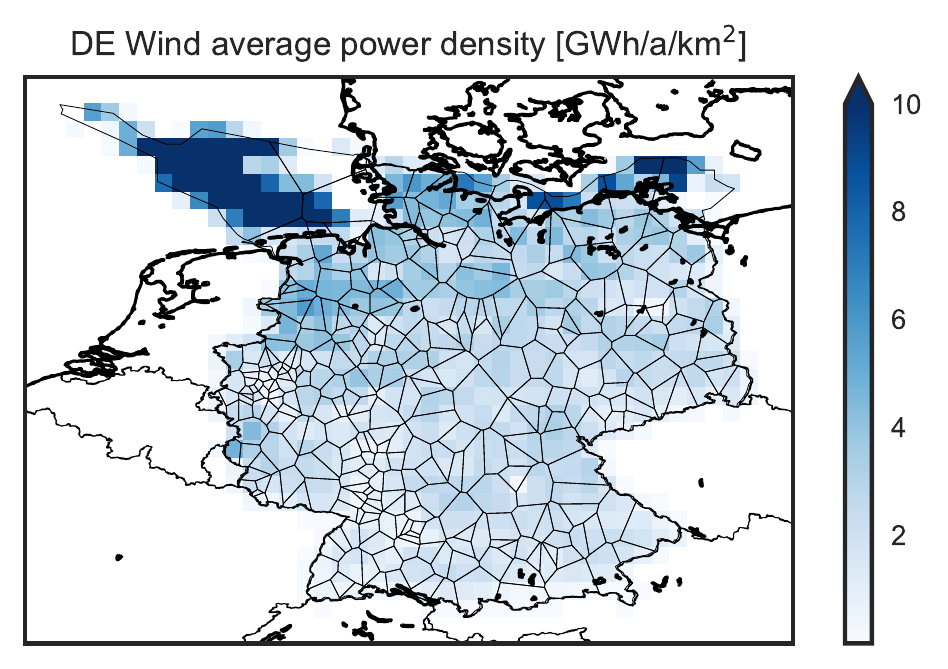}
\includegraphics[width=0.48\textwidth]{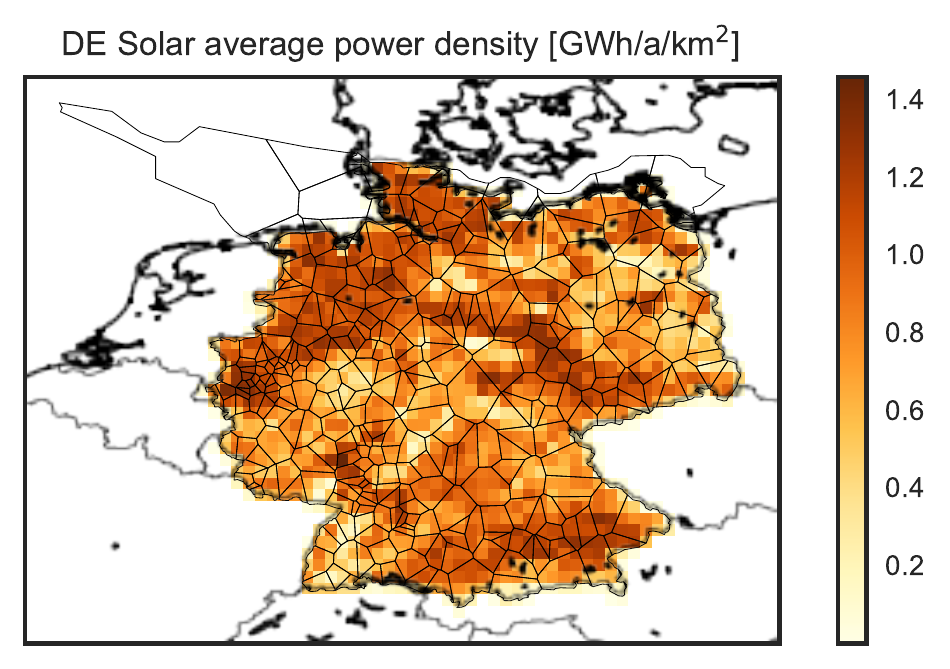}
\end{figure}

The solar availability time-series and maximally installable capacity per substation are based on the direct and diffuse surface solar irradiance in Surface Solar Radiation Data Set - Heliosat (SARAH-2), but except for that is similar to the generation time-series of wind turbines.

The photovoltaic generation $P_{x,s}(t)$ for a panel of nominal capacity $P^{max}_{s}$ of a point in time $t$ and space resp. grid cell $x$ is calculated from the surface solar irradiance. The total panel irradiation is derived from the solar azimuth and altitude~\cite{michalsky1988} using geometric relations of the trajectory of the sun and the tilted panel surface~\cite{sproul2007}. All solar panels are facing South at an angle of \SI{35}{\deg}. The electric model by Huld~et~al.~\cite{huld2010} determines the active power output from the total irradiation and the ambient temperature. Implementation details are found in the \texttt{pv} sub-package of the atlite package \cite{atlite-github}.

For each raster cell $x \in \mathcal{X}$ the capacity factor $c_{x,s}$ and the maximally installable capacity $G^{max}_{x,s} = \num{0.01} \cdot \SI{145}{MW/km^2} \cdot A_{x,s}$ is determined as for wind, with the difference that the high technical potential of \SI{145}{MW/km^2} corresponds to an unrealistic full surface of solar cells, which is offset by allowing only up to $1\%$. The permitted CORINE land use types are \emph{Artificial surfaces}, most \emph{Agricultural areas} except for those with forests and then including only few sub-categories of \emph{Forest and semi natural areas}: \emph{Scrub and/or herbaceous vegetation associations}, \emph{Bare rocks} and \emph{Sparsely vegetated areas}. Figure~\ref{fig:potentials} shows the solar generation potentials.

Equations~(\ref{eq:wind-caplayout})-(\ref{eq:wind-capcap}) are applied analogously to generate the solar availability time-series $\bar g_{V,s}(t)$ and to find the solar expansion potential $G^{max}_{V,s}$. The reference solar panel is the crystalline sillicon panel fitted in \cite{huld2010}.

\subsection{Demand}

The hourly electricity demand profiles for each country from 2011 to 2016 are taken from the European Network of Transmission System Operators for Electricity (ENTSO-E) website~\cite{entsoe_load}. The load time-series is distributed to the substations in each country by 60\% according to the gross domestic product (GDP) as a proxy for industrial demand and by 40\% as residential demand according to population in a Voronoi cell. The 60-40\% split is based on a linear regression analysis of the per-country data and agrees with values used in~\cite{ELMOD-DE}. The two statistics are mapped from the Eurostat Regional Economic Accounts database (nama\_10-reg) for NUTS3 regions to the Voronoi cells in proportion to their geographic overlap.

\section{Validation}
\label{sec:validation}

\subsection{Network total line lengths}

In this subsection, total line circuit lengths at different voltage levels in the model are compared with official statistics from ENTSO-E.  The lengths of AC circuits~\cite{entsoe-lengths-of-circuits} per voltage level and country are compared to aggregations of line lengths times circuits from PyPSA-Eur, so that cross-border lines are equally attributed to both adjacent countries. In Table~\ref{tab:lengths-of-circuits} the total line lengths for the whole of Europe and Germany are presented as examples. Considering the data for all countries, the lines in the PyPSA-Eur dataset deviate from the ENTSO-E lengths of circuits by a mean absolute error of 15\% for \SI{220}{kV}, 7\% for \SI{300}{kV} and 9\% for \SI{380}{kV} lines. These deviations are accounted for by the fact that the ENTSO-E map \cite{interactive} from which the PyPSA-Eur network is derived is only an artistic representation and does not follow the exact contours of each transmission line. Some differences may also be due to incorrect classification of 220~kV lines as 380~kV lines, or due to the fact that the ENTSO-E map on which  PyPSA-Eur is based is more up-to-date with regard to recent upgrades to the transmission network.

\begin{table}
  \caption{AC lines circuit lengths of the whole of Europe and Germany as an example}
  \label{tab:lengths-of-circuits}
\begin{tabularx}{\linewidth}{X SSS SSS}
  \toprule
  Circuit length & \multicolumn{3}{c}{DE}            &  \multicolumn{3}{c}{EU} \\
  in \SI{1000}{km} & \SI{220}{kV} & \SI{300}{kV} &      \SI{380}{kV} &  \SI{220}{kV} &     \SI{300}{kV} &       \SI{380}{kV} \\
  \midrule
  ENTSO-E   &  13.70 &   0.0 &  20.92 &   117.25 &  9.96 &  146.82 \\
  PyPSA-Eur &  10.49 &   0.0 &  24.97 &   116.69 &  9.23 &  154.31 \\
  \bottomrule
\end{tabularx}

\end{table}

\subsection{Network topology}

While the total circuit lengths might agree, it does not necessarily mean that the lines are in the right places with the right topology. The \emph{ENTSO-E Interconnected network map}~\cite{interactive} is the source of the network topology in PyPSA-Eur, so it naturally agrees well in visual examination. A comparison with the network topology published with the TYNDP~\cite{STUM} is hindered by shortened substation names and missing geo-locations in that dataset. Instead, in this section the network topology of PyPSA-Eur is compared to the open network datasets available for Germany, which are derived using a different methodology. New, experimental algorithms are presented to compare network topologies, since few appropriate algorithms exist in the literature. This is a difficult problem because neither the locations nor the number of the buses and lines in the different models necessarily agree. Our methodology works by first establishing a common set of aggregated buses for the different networks, then comparing the networks once all lines and other elements have been reattached to the aggregated buses.

\begin{figure}
\centering
\includegraphics[width=\textwidth]{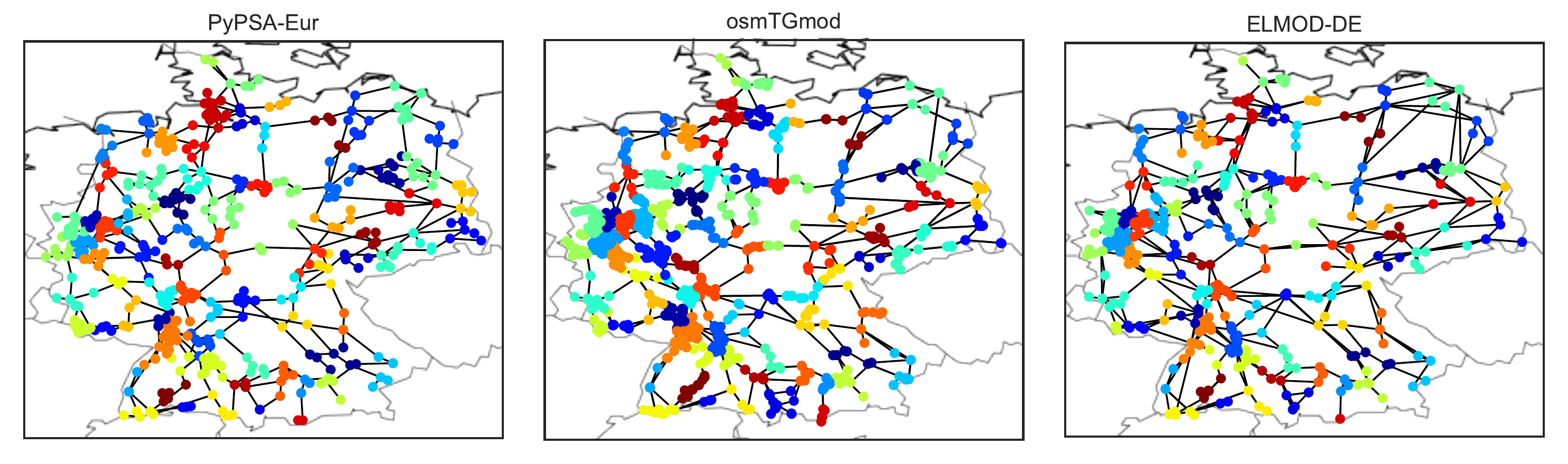}
\caption{\num{80} clusters jointly identified by colour in the network topologies of
  the models PyPSA-Eur, osmTGmod and ELMOD-DE.}
\label{fig:validation-clusters}
\end{figure}

More precisely, we present a new technique of applying $k$-means clustering to measure the similarity of several geo-located network models, specifically the German \SI{220}{kV} and \SI{380}{kV} voltage layers represented in PyPSA-Eur, osmTGmod and ELMOD-DE. The buses of all networks are jointly clustered together into $k$ clusters by minimizing the distances in each cluster $\pi_j$
\begin{linenomath}
\begin{equation}
  \mathcal{D}({\pi_j}) = \sum_{j=1}^k \sum_{a \in \pi_j} w(a) \,||a - m_j||^2
  \label{eq:kernel-kmeans}
\end{equation}
\end{linenomath}
from its buses $a$ to the center $m_j = \sum_{b \in \pi_j} w(b) \, b / \sum_{b \in \pi_j} w(b)$ with chosen bus-weights $w(b)$. Since Eq.~(\ref{eq:kernel-kmeans}) expands to a polynomial in the scalar products $\left\langle\, a, b \,\right\rangle$ between buses, kernel k-means allows the use of general scalar products by evaluating them on every pair of buses~\cite{dhillon2004}. In the kernel of the scalar product we propose
\begin{linenomath}
\begin{equation}
K_{a,b} = e^{- ||a-b||_2^2/N} + \nu B^{\dag}_{a,b}
\end{equation}
\end{linenomath}
spatial cohesion comes from the first term, a radial basis function based on the Euclidean distance $||\cdot||_2$ over the number of buses $N$, which favours short geometric distances and connectedness over spatial convexity, while the pseudo inverse of the admittance matrix $B$ in each network induces an electrical reactance distance, shown to lead to electrically cohesive clusters~\cite{cotilla-sanchez2013}. With the weights $w(a \in \text{PyPSA-Eur}/\text{ELMOD-DE}) = 5$ and $w(a \in \text{osmTGmod})=1$ balancing the five times as many buses in osmTGmod and the relative weight $\nu = 200$, the clustering algorithm is able to distribute \num{80} clusters across the three networks, by starting from the labels found by regular $k$-means and by picking the best result from 20 runs. The networks and the associated buses are shown in Figure~\ref{fig:validation-clusters}.

\begin{figure}
\centering
\includegraphics[width=0.9\textwidth]{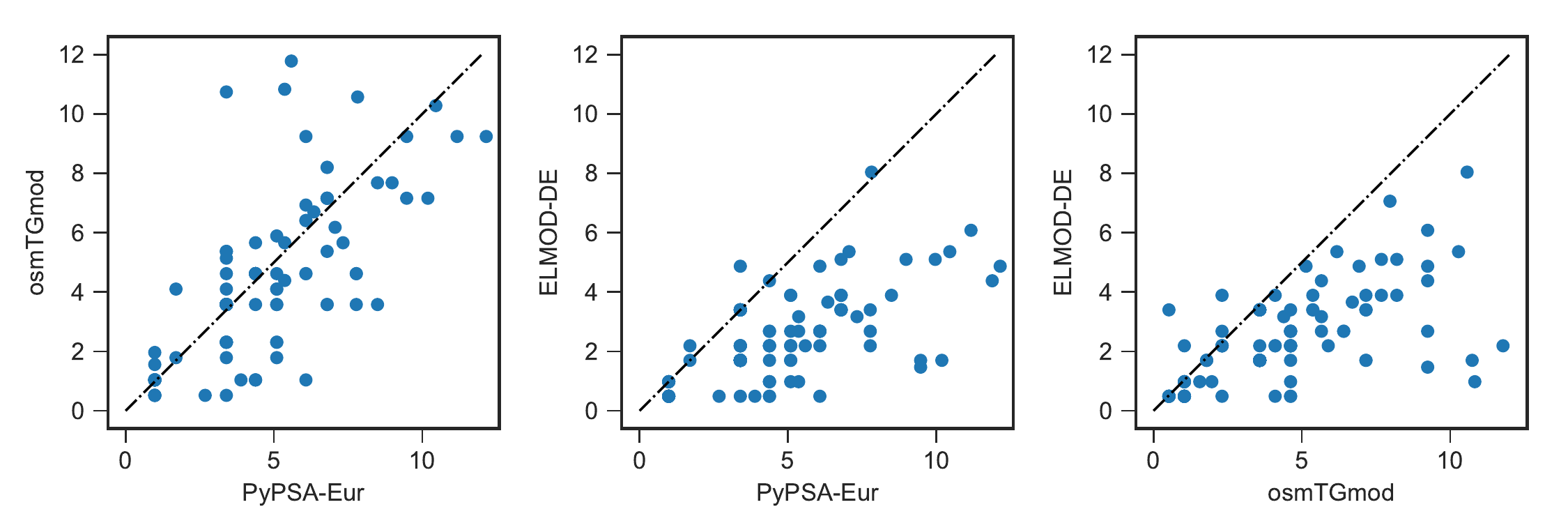}
\caption{Capacity connecting the same clusters in GW}
\label{fig:validation-inter-cluster-capacity}
\end{figure}

As the clusters are the same in each network, the aggregate capacity between every two clusters is now comparable for different networks. Unfortunately, Figure~\ref{fig:validation-inter-cluster-capacity} reveals bad agreement for the capacity between pairs of clusters, due to its high sensitivity to errors arising from clustering topologically distinct buses; i.e. buses lying on distinct lines are inadvertently joined together. Increasing the weight of the electrical distance $\nu$ dampens the appearance of these associations, but worsens convergence and increasingly finds solutions in which clusters in the electrically well connected areas as the Ruhrpott detach from one of the three networks. With a lower number of clusters between $20$ to $40$, $\nu$ can be increased by an order of magnitude and the topological errors are less important, then the aggregate capacities between large network zones can be compared and deviations identified.

\begin{figure}
\centering
\includegraphics[width=0.9\textwidth]{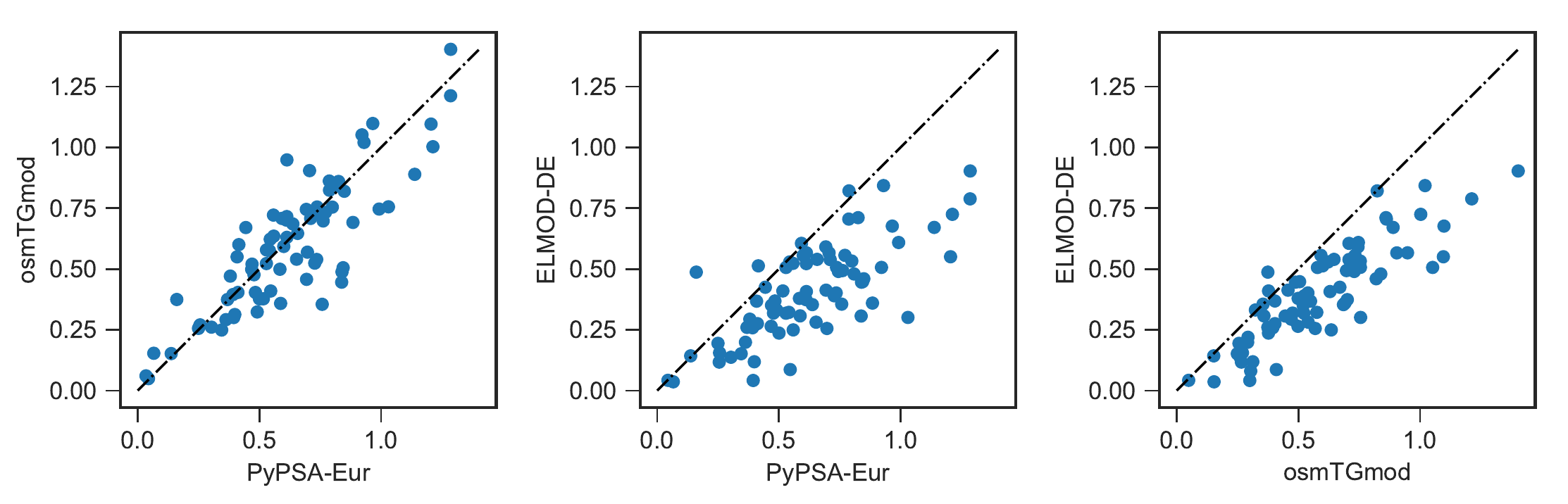}
\caption{Line volume at buses in the same cluster in TWkm}
\label{fig:validation-lv-corr}
\end{figure}

Another approach is simply to aggregate line volumes within and attached to each cluster of buses. These are compared in Figure~\ref{fig:validation-lv-corr} and turn out to be quite robust against topologically problematic associations and show a high correlation across networks in Table~\ref{tab:validation-lv-corr}. ELMOD-DE has proportionally less line volume than osmTGmod and PyPSA-Eur, but with approximately the same spatial distribution. osmTGmod and PyPSA-Eur agree well.

\begin{table}
\centering
\caption{Pearson correlation coefficients between line volume at different buses}
\label{tab:validation-lv-corr}
\begin{tabular}{lrrr}
\toprule
{} &  PyPSA-Eur &  osmTGmod &  ELMOD-DE \\
\midrule
PyPSA-Eur &   1.000 &  0.856 &  0.741 \\
osmTGmod  &    &  1.000 &  0.868 \\
ELMOD-DE  &    &   &  1.000 \\
\bottomrule
\end{tabular}
\end{table}







\subsection{Potentials for expansion of renewables}

Geographic potentials for the expansion of wind and solar power depend strongly on technical, environmental, social and political constraints. Different organisations offer different assessments of acceptable potentials, which involve a complex balance between land availability, landscape impact and species protection. In this section we compare aggregated total potentials for Germany in the PyPSA-Eur model derived using the methodologies described in Sections \ref{sec:wind} and \ref{sec:PV} with other studies.

For onshore wind, there is an installable potential of \SI{441}{GW} in Germany in the model. Assessments in the literature range from \SI{198}{GW}~\cite{bwe2012} (based on a `realistic' restriction to 2\% of total land area, although 8\% is available when excluding forests and protected areas) up to \SI{1190}{GW}~\cite{ubapot2013} (using 13.8\% of the total land area, ignoring species protection and whether locations are economically exploitable).

For offshore wind, \SI{87}{GW} of fixed-foundation capacity is installable in PyPSA-Eur in Germany. Estimates in the literature range from \SI{38}{GW}~\cite{IEESWV} to \SI{85}{GW}~\cite{ISE2012}.


\SI{350}{GW} of solar photovoltaics is installable in Germany in PyPSA-Eur. The potential depends strongly on what land areas are permitted, but typical values range from \SI{360}{GW}~\cite{IEESWV} to \SI{400}{GW}~\cite{ISE2012} (including roofs, facades and railway/motorway sidings, but excluding free space).

\subsection{Model validation: Linear optimal power flow}

\begin{figure}
\floatbox[{\capbeside\thisfloatsetup{capbesideposition={left,bottom},capbesidewidth=0.3\linewidth}}]{figure}[\FBwidth]
{\caption{Load shedding and line loading at peak demand.}\label{fig:loadshedding-map}}
{\includegraphics[width=0.6\textwidth]{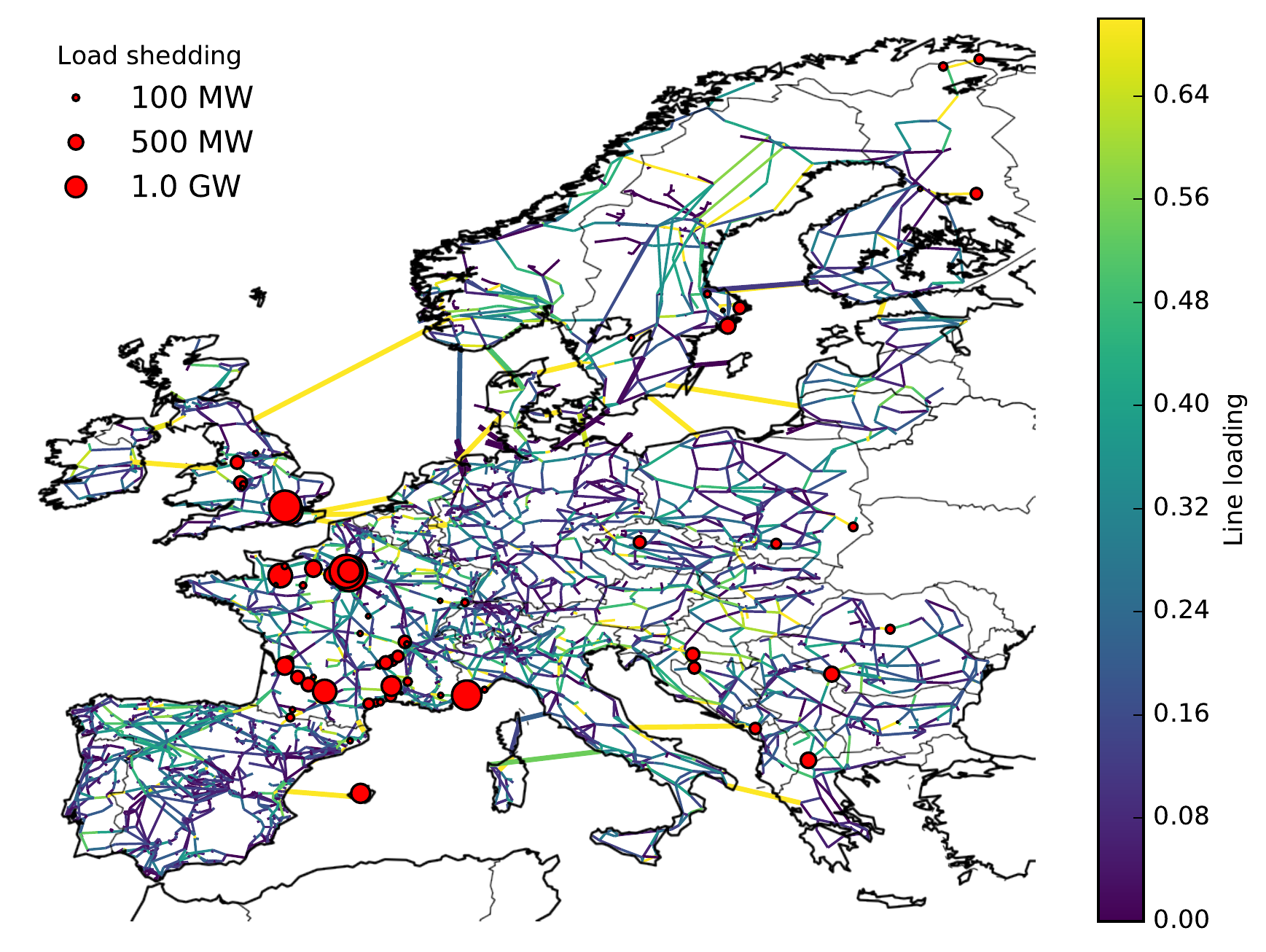}}
\end{figure}

As a validation of the model at large, it is formulated within the Python for Power System Analysis (PyPSA) framework~\cite{pypsa-0.13.1,pypsa} and the linear optimal power flow of the European peak-load hour is considered to check the feasibility of the combined network, generation and demand data for supplying the most extreme demand, so that line-loading is below 70\% to approximate the $N-1$ stability constraint.

Solar- and wind feed-in are not allowed to reduce the load, while hydro-electric installations may be discharged at their full power capacity. European overall peak-load of \SI{0.51}{TW} in the dataset happens at 17:00 on 17-01-2013 and leads to \SI{20.4}{GW} of load shedding in the vicinity of large agglomerations, primarily in Paris (\SI{6.2}{GW}) and London (\SI{3.8}{GW}) as shown in Fig.~\ref{fig:loadshedding-map}. Since there is sufficient generation capacity to cover the peak load, this load-shedding is due to grid bottlenecks which appear in the model (but not in reality, since grid bottlenecks do not cause load-shedding in today's European network).
The amount of shedding decreases considerably by lifting capacity constraints on short lines, for example if lines shorter than \SI{25}{km} are not limited in power capacity, only about \SI{6}{GW}, 1\%, of load has to be shed. Similarly easing local restrictions by clustering the network using a k-means algorithm as detailed in~\cite{hoersch2017-eem} to \num{1500} buses reduces shedding to 1\% of peak load, while clustering to \num{362} buses allows most of the demand to be supplied, except for \SI{650}{MW} on Mallorca, where the power plant dataset is missing \SI{1.4}{GW} of coal and gas capacities, and at the Northern tip in Norway. This is indicative of local assignment errors of load and supply, when the Voronoi cells used to assign load and generators to transmission substations do not represent the true distribution grid topology at each transmission substation, and/or an underrepresentation of inner-city underground cabling, which is not always shown on the map. Clustering the network, so that each bus represents a larger area, smooths out local assignment errors. Since there are several heuristic remedies from expanding the loaded lines over rearranging the load to using clustered topologies as done manually by \cite{Zhou2005,Hutc13}, we decided not to perform any corrections, but publish the dataset as is.

\section{Limitations}
\label{sec:limitations}

While the benefit of an openly available, functional and partially validated model of the European transmission system is high, many approximations have been made due to missing data. In this section we summarise the limitations of the dataset, both as a warning to the user and as an encouragement to assist in improving the approximations.

The grid data is based on a map of the ENTSO-E area \cite{interactive} that is known to contain small distortions to improve readability. Since the exact impedances of the lines are unknown, approximations based on line lengths and standard line parameters were made that ignore specific conductoring choices for particular lines. There is no openly available data on busbar configurations, switch locations, transformers or reactive power compensation assets.

Using Voronoi cells to aggregate load and generator data to transmission network substations ignores the topology of the underlying distribution network, meaning that assets may be connected to the wrong substation. Assumptions have been made about the distribution of load in each country proportional to population and GDP that may not reflect local circumstances. Openly available data on load time series may not correspond to the true vertical load \cite{schumacher2015} and is not spatially disaggregated; assuming, as we have done, that the load time series shape is the same at each node within each country ignores local differences.

Information on existing wind, solar and small hydro, geothermal, marine and biomass power plants are excluded from the dataset because of a lack of data availability in many countries. Approximate distributions of wind and solar plants in each country can be generated that are proportional to the capacity factor at each location.

The database of hydro-electric power plants does not include plant-specific energy storage information, so that blanket values based on country storage totals have been used. Inflow time series are based on country-wide approximations, ignoring local topography and basin drainage; in principle a full hydrological model should be used. Border connections and power flows to Russia, Belarus, Ukraine, Turkey and Morocco have not been taken into account; islands which are not connected to the main European system, such as Malta, Crete and Cyprus, are also excluded from the model.

\section{Conclusions}
\label{sec:conclusions}

In this paper a dataset PyPSA-Eur has been presented of the full European transmission system, including a high resolution grid model, load data, a new geo-referenced database of conventional power plants, potentials for the expansion of wind and solar, and time series for the load and variable renewable power availability. The model is only based on publicly available and open datasets, and all code and data has been made available~\cite{pypsa-eur-zenodo,pypsa-eur-github}, making it the first open model of the full European system at such high spatial resolution.

To validate the model, total circuit lengths were compared with official statistics for Europe, renewable expansion potentials were checked against literature values, an optimal power flow study was performed, and a new technique was developed to compare the network topology with other network models. Together these validation steps demonstrate that the model is a plausible approximation of the European power system. Further validation is desirable to increase confidence in the model.

Since PyPSA-Eur is open, it can be further improved by any research group as and when better data or new methodologies become available. It is also hoped that the existence of unofficial open datasets such as PyPSA-Eur will encourage data holders to release their own official datasets, in the interests of improving modelling by third parties.

Given the pressing need to understand how to adapt the electricity system to rising shares of variable renewable generators, new market structures, the electrification of other energy sectors such as transport and heating, and rising public concerns about the landscape impacts of overhead lines, there is a clear imperative for detailed, rigorous and reproducible modelling of the transmission system. The dataset PyPSA-Eur has been designed primarily for the optimisation of future investment in generation and transmission, but can also be adapted to studies of the operation of the current power system. We hope that it will contribute towards a transparent discussion of the future needs of the European energy system.

\section*{Acknowledgements}

All authors acknowledge funding from the German Federal Ministry of
Education and Research under grant no. 03SF0472C. T.B. and J.H.
acknowledge funding from the Helmholtz Association under grant no.
VH-NG-1352.

\bibliography{pypsa-eur}

\end{document}